\documentclass[aps, prb, twocolumn, showpacs, amsmath, amssymb]{revtex4}
\usepackage{graphicx}
\usepackage{bm}

\begin{document}
\bibliographystyle{apsrev}

\title{Slow dynamics at the smeared phase transition of randomly layered magnets}
\author{Shellie Huether}
\author{Ryan Kinney}
\author{Thomas Vojta}
\affiliation{Department of Physics, University of Missouri -- Rolla, Rolla, MO 65409, USA}
\date{\today}

\begin{abstract}
We investigate a model for randomly layered magnets, viz.\ a three-dimensional Ising model with
planar defects. The magnetic phase transition in this system is smeared because static long-range
order can develop on isolated rare spatial regions.  Here, we report large-scale kinetic Monte Carlo
simulations of the dynamical behavior close to the smeared phase transition which we characterize
by the spin (time) autocorrelation function. In the paramagnetic phase, its behavior is dominated by Griffiths
effects similar to those in magnets with point defects. In the tail region of the smeared
transition the dynamics is even slower: the autocorrelation function decays like a stretched
exponential at intermediate times before approaching the exponentially small asymptotic value
following a power law at late times. Our Monte-Carlo results are in good agreement with
recent theoretical predictions based on optimal fluctuation theory.
\end{abstract}


\maketitle

\section{Introduction}
\label{sec:Int}

In recent years, there has been a resurgent interest in the influence of defects,
impurities or other types of quenched disorder on the properties of phase transitions and
critical phenomena. This renewed attention is largely due to the discoveries of novel
disorder effects that go beyond the framework of perturbation theory and the perturbative
renormalization group.

A particularly interesting class of non-perturbative phenomena are the so-called  Griffiths
or rare region effects that are produced by rare strong spatial disorder
fluctuations. They can be easily understood on the example of a diluted magnet: Due to
the dilution, the critical temperature $T_c$ of the disordered system is lower than its
clean value, $T_c^0$. In the temperature interval $T_c<T<T_c^0$, the diluted bulk system
is in the disordered phase. However, in an infinite size sample, there is an
exponentially small, but nonzero probability for finding an arbitrary large spatial region devoid
of impurities. Such a rare region, a 'Griffiths island', displays local order in the
temperature interval $T_c<T<T_c^0$. Due to its size, such an island will have very slow
dynamics because flipping it requires changing the order parameter over a large
volume. More than 30 years ago, Griffiths \cite{Griffiths69} showed that the presence of
these locally ordered islands produces a singularity in the free energy in the whole
region $T_c<T<T_c^0$, which is now known as the Griffiths region or the Griffiths
phase.\cite{RanderiaSethnaPalmer85} In generic classical systems with uncorrelated or
short-range correlated disorder, thermodynamic Griffiths effects are very weak because
the singularity in the free energy is only an
essential one. \cite{Wortis74,Harris75,BrayHuifang89} To the best of our knowledge,
classical thermodynamic Griffiths singularities have therefore not been verified in
experiment (see also Ref.\ \onlinecite{Imry77}).

In contrast to the thermodynamics, the long-time \emph{dynamics} is dominated by the rare
regions. Inside the Griffiths phase, the spin autocorrelation function $C(t)$ decays very
slowly with time $t$, as $\ln C(t)\sim -(\ln t)^{d/(d-1)}$ for Ising systems
\cite{Dhar83,RanderiaSethnaPalmer85,DharRanderiaSethna88,Bray88,BrayRodgers88} and as
$\ln C(t)\sim -t^{1/2}$ for Heisenberg systems.\cite{Bray88,Bray87} These results were
recently confirmed by more rigorous calculation for the equilibrium
\cite{DreyfusKleinPerez95,GielisMaes95}
and dynamic\cite{CesiMaesMartinelli97a,CesiMaesMartinelli97b} properties of disordered Ising systems.

The effects of impurities and defects are greatly enhanced by long-range spatial disorder
correlations. In particular, if the disorder is perfectly correlated in some spatial directions,
the rare regions are extended objects which are infinite in the correlated dimensions.
This makes their dynamics even slower and so increases their effects. In Ising models with
linear defects (disorder perfectly correlated in one dimension), the thermodynamic Griffiths
singularities are of power-law type, with the average susceptibility actually diverging
in a finite temperature region. The critical point itself is an exotic
infinite-randomness critical point and displays activated rather than power-law scaling. This was
first found in the McCoy-Wu model, \cite{McCoyWu68,McCoyWu68a} a disordered 2D Ising model in
which the disorder is perfectly correlated in one dimension and uncorrelated in the
other. Later it was studied in great detail in the context of the quantum phase
transition of the random transverse-field Ising model where the imaginary time dimension
plays the role of the ``correlated'' direction. \cite{Fisher92,Fisher95}

Recently, it has been shown that even stronger effects than these power-law Griffiths singularities
can occur in Ising models with planar defects. \cite{Vojta03b,SknepnekVojta04} Because the disorder
is perfectly correlated in two directions, the effective dimensionality of the rare regions is two.
Therefore, an isolated rare region can undergo the phase transition independently from the bulk system.
This leads to a destruction of the global sharp phase transition by smearing. Similar smearing effects
have also been found in itinerant quantum magnets \cite{Vojta03a} and at a non-equilibrium transition
in the presence of linear defects \cite{Vojta04}. A recent review of these and other rare
region effects can be found in Ref.\ \onlinecite{Vojta06}.

In this paper, we study  the {\em dynamics} of an Ising model with planar defects in
the vicinity of this smeared phase transition by large-scale kinetic Monte-Carlo
simulations. The paper is organized as follows. In Section \ref{sec:theory} we introduce
the model and briefly summarize the results of the optimal fluctuation theory
\cite{Vojta03b,FendlerSknepnekVojta05,Vojta06} for smeared phase transitions to the extent
necessary for the analysis of our data. In section \ref{sec:simulations}, we explain
the simulation technique, we present our results for the spin autocorrelation function,
and we compare them to the theoretical predictions. Conclusions are presented in section
\ref{sec:conclusions}.

\section{Theory}
\label{sec:theory}
\subsection{Three-dimensional Ising model with planar defects}
\label{subsec:model}

We consider a three-dimensional classical Ising ferromagnet with planar defects, the same
model whose thermodynamics was studied in Ref.\ \onlinecite{SknepnekVojta04}.
Ising variables $S_{i,j,k}=\pm 1$ reside on the sites of a cubic lattice. They interact via nearest-neighbor
interactions. In the clean system all interactions are identical and have the value $J$.  The defects are
modeled via 'weak' bonds randomly distributed in one dimension (uncorrelated direction). The bonds
in the remaining two dimensions (correlated directions) remain equal to $J$. The system
effectively consists of strongly-coupled slabs (layers) of varying thickness, separated by parallel layers
of weak bonds. The Hamiltonian of the system is given by:
\begin{eqnarray}
 H=&&-\sum_{ {i=1,\dots,L_\bot} \atop
{j,k=1,\dots,L_C} }J_iS_{i,j,k}S_{i+1,j,k}\nonumber\\ && - \sum_{{i=1,\dots,L_\bot} \atop
{j,k=1,\dots,L_C}}J(S_{i,j,k}S_{i,j+1,k}+S_{i,j,k}S_{i,j,k+1}),
\label{eq:Hamiltonian}
\end{eqnarray}
where $L_\bot$($L_C$) is the length in the uncorrelated (correlated) direction, $i$, $j$ and $k$
are integers counting the sites of the cubic lattice, $J$ is the interaction energy in the
correlated directions and $J_i$ is the random interaction in the uncorrelated direction. The
$J_i$ are drawn from a binary distribution:
\begin{equation}
\label{eq:distrib} J_i=\left\{
\begin{array}{clc}
 cJ & \textrm{with probability} & p\\ J  & \textrm{with probability} & 1-p
\end{array}
\right .
\end{equation}
characterized by the concentration $p$ with $0\le p<1$ and the relative strength $c$ of the weak bonds ($0<c\le
1$).

The Hamiltonian (\ref{eq:Hamiltonian}) does not contain any internal dynamics. Instead,
the dynamics must be added ad-hoc by specifying an appropriate dynamical algorithm. We
consider a purely relaxational local dynamics without any conservation laws, i.e., model
A in the Hohenberg-Halperin classification.\cite{HohenbergHalperin77} Microscopically, it
can be realized, e.g., by the Glauber or Metropolis algorithms.\cite{MRRT53,Glauber63}

\subsection{Smeared phase transition and optimal fluctuation theory}
\label{subsec:OFT}

In the absence of impurities ($p=0$), the system (\ref{eq:Hamiltonian}) undergoes a
ferromagnetic phase transition at the clean critical temperature $T_c^0=4.511 J$, with the order
parameter being the total magnetization:
\begin{equation}
\label{eq:magnetization} m=\frac{1}{V} \sum_{i,j,k} \langle S_{i,j,k} \rangle~.
\end{equation}
Here, $V=L_\bot L_C^2$ is the volume of the system, and $\langle\cdot\rangle$ denotes the thermodynamic
average.

In the presence of disorder ($p>0$), a crucial role is played by rare strong disorder
fluctuations: In analogy to the Griffiths phenomena\cite{Griffiths69} discussed in the Introduction,
there is a small but finite probability for finding
a large spatial region containing only strong bonds in the uncorrelated direction. Such a rare region
can be locally in the ferromagnetic state while the bulk system is still in the disordered (paramagnetic) phase. The
ferromagnetic order on the largest rare regions starts to emerge right below the clean critical
temperature $T_c^0$. Since the defects in the system are planar, these rare regions are
infinite in the two correlated dimensions but finite in the uncorrelated direction. Each rare region
is thus equivalent to a two-dimensional Ising system that can undergo a real phase transition
\emph{independently} of the rest of the system. The resulting effect is much stronger than conventional
Griffiths effects: the global phase transition is destroyed by smearing, and the order parameter develops
very inhomogeneously in space with different parts of the system (different rare regions)
ordering independently at different temperatures.\cite{Vojta03b,SknepnekVojta04,Vojta06}

The leading thermodynamic behavior in the tail of the smeared transition can be
determined using optimal fluctuation theory.\cite{Vojta03b,Vojta06} The approach is similar
to that of Lifshitz \cite{Lifshitz64} and others for the description of the tails in
the electronic density of states of disordered systems. The theory can be easily developed for a general $d$-dimensional
system with $d_C$ correlated dimensions and $d_\perp=d-d_C$ uncorrelated dimensions. In the case of the Hamiltonian
(\ref{eq:Hamiltonian}), $d_C=2$ and $d_\perp=1$.

In the tail region of the smeared transition, the system consists of a few isolated ferromagnetic rare regions
embedded in a paramagnetic bulk. We start from the probability $w$ for finding a large region of linear
size $L_{RR}$ containing only strong bonds. Up to pre-exponential factors, it is given
by
\begin{equation}
\label{eq:w} w\sim (1-p)^{L_{RR}^{d_\perp}} = \exp \left[ {L_{RR}^{d_\perp}\ln(1-p)}\right] .
\end{equation}
Such a rare region develops static ferromagnetic long-range order at some temperature $T_c(L_{RR})$
below the clean critical temperature $T_c^0$. The value of $T_c(L_{RR})$ varies with the size of the rare
region; the largest islands will develop long-rage order closest to the clean critical point. To determine
$T_c(L_{RR})$, we can use finite size scaling \cite{Barber_review83} for the clean system because a
rare region is equivalent to a ``slab'' of a clean Ising model. This yields
\begin{equation}
\label{eq:FSS} T_c^0-T_c(L_{RR})=|r_c(L_{RR})|=AL_{RR}^{-\phi},
\end{equation}
where $\phi$ is the finite-size scaling shift exponent of the clean system and A is the
amplitude for the crossover from $d=d_C+d_\perp$ dimensions to a slab geometry infinite
in $d_C$ (correlated) dimension but with finite length in the other (uncorrelated)
directions. The reduced temperature $r = T - T_c^0$ measures the distance from the
\emph{clean} critical point. As long as the clean $d$-dimensional Ising model is below
its upper critical dimension ($d_c^{+}=4$), hyperscaling is valid and the finite-size
shift exponent is related to the correlation length exponent $\nu$ by $\phi=1/\nu$ which
we assume from now on. Combining (\ref{eq:w}) and (\ref{eq:FSS}), we obtain the
probability for finding an island  which becomes critical at some $r_c$ as:
\begin{equation}
\label{eq:w(r)}
w(r_c)\sim \exp \left({-B|r_c|^{-d_\perp\nu}}\right) \quad \textrm{(for $r_c\to 0-$)}
\end{equation}
with the constant $B=-\ln(1-p)A^{d_\perp \nu}$. The total (average) magnetization $m$ at some reduced
temperature $r$ is obtained by integrating over all rare regions which are ordered at $r$, i.e., those with
$r_c>r$. Since the functional dependence on $r$ of the local magnetization on the island is of power-law type it does
not enter the leading exponentials but only pre-exponential factors. To
exponential accuracy, we therefore obtain
\begin{equation}
\label{eq:mtot}
m(r)\sim \exp \left({-B|r|^{-d_\perp\nu}}\right) \quad \textrm{(for $r\to0-$)}.
\end{equation}
Thus, the total
magnetization develops an exponential tail towards the disordered phase which reaches all
the way to the clean critical point. Analogous estimates show that the homogeneous magnetic
susceptibility does not diverge anywhere in the tail region of the smeared transition.
At $r=0$, the exponentially decreasing island density overcomes the power-law divergence of
the susceptibility of an individual island; and once a nonzero magnetization has
developed, it cuts off any possible singularity.
 However, there is an essential singularity in the free energy at
the clean critical point produced by the vanishing density of ordered islands.

\subsection{Dynamics at the smeared transition}
\label{subsec:OFT-dynamics}

After having briefly discussed the optimal fluctuation theory for the thermodynamics, we
now consider the dynamical behavior at the smeared phase transition of our disordered
magnet (\ref{eq:Hamiltonian}). The interesting physics in the tail of the smeared transition
is local with respect to the uncorrelated dimensions because different rare regions
are effectively decoupled from each other, and the spatial correlation length
in these directions remains finite. An appropriate quantity to study the rare region dynamics is therefore
the time autocorrelation function of the Ising spins,
\begin{equation}
C(t) = \frac 1 V \sum_{i,j,k} ~\langle S_{i,j,k}(t) S_{i,j,k}(0)\rangle~.
\label{eq:autocorr}
\end{equation}
The leading long-time behavior of $C(t)$ in the tail of the smeared transition can be determined
using optimal fluctuation arguments similar to that of section
\ref{subsec:OFT}.\cite{Vojta06,FendlerSknepnekVojta05}

According to finite-size scaling \cite{Barber_review83}, the behavior of the correlation time
$\xi_t$ of a single rare region of size $L_{RR}$ in the vicinity of the clean bulk critical
point can be modelled by (for $T<T_c^0$, i.e., $r<0$)
\begin{equation}
\xi_t(r, L_{RR}) \sim L_{RR}^{(z\nu - \tilde z\tilde \nu)/\nu}
   \left|  r + A L_{RR} ^{-1/\nu}\right|^{-\tilde z \tilde \nu}~. \label{eq:xit}
\end{equation}
Here, $\nu$ and $z$ are the correlation length and dynamical exponents of a $d$-dimensional system, and $\tilde \nu$ and
$\tilde z$ are the corresponding exponents of a $d_C$-dimensional system.

Let us first consider the time evolution of the autocorrelation function $C(t)$ at the
clean critical point $T_c^0$, i.e., the boundary between the conventional paramagnetic phase
and the smeared tail of the ordered phase. For $r=0$, the correlation time (\ref{eq:xit}) simplifies
to $\xi_t \sim L_{RR}^z$. The rare region contribution to $C(t)$ is obtained by simply
summing over the exponential time dependencies of the individual islands with
$L_{RR}$-dependent correlation time. Using (\ref{eq:w}) we obtain to exponential accuracy
\begin{equation} C(t) \sim \int dL_{RR}
~\exp \left ( L_{RR}^{d_\perp} \ln(1-p) -D t/L_{RR}^z \right)
\label{eq:ct-oft}
\end{equation}
where $D$ is a constant. This integral can be easily estimated within the saddle-point
method.  The leading long-time decay of the autocorrelation function $C(t)$ at the clean
critical point is given by a stretched exponential,
\begin{equation}
\ln C(t) \sim -[-\ln(1-p)]^{z/(d_\bot+z)}~ t^{d_\bot/(d_\bot+z)} \qquad (r=0). \label{eq:stretched}
\end{equation}

In the conventional paramagnetic phase, $T>T_c^0$ ($r>0$), the correlation time does not diverge
for any $L_{RR}$. Instead, the correlation time of the large islands saturates at
$\xi_t(r, L_{RR}) \sim r^{-z \nu}$ for $L_{RR} > (r/A')^{-\nu}$. The autocorrelation function
$C(t)$ can again be evaluated as an integral over all island contributions. We find that
there are two regimes separated by a crossover time
\begin{equation}
t_x\sim |r|^{-(d_\bot+z)\nu}~.
\end{equation}
For intermediate times $t<t_x$, the autocorrelation function follows the stretched exponential
(\ref{eq:stretched}). For times larger than the crossover time $t_x$, we obtain a simple exponential
decay
\begin{equation}
\ln C(t) \sim -t/\tau  \qquad (\textrm{for}~ t>t_x, ~r>0)
\label{eq:simple}
\end{equation}
with the decay time $\tau \sim r^{-z\nu}$. Our
results for $T \ge T_c^0$ agree with the corresponding conventional dynamical Griffiths
effects\cite{Bray88} for systems with point defects. This is not surprising, because above
$T_c^0$, there is no qualitative difference between the Griffiths and the smearing scenarios: In both cases,
all rare regions are locally still in the disordered phase.

This changes in the tail of the smeared transition below the clean critical point $T_c^0$.
For $r<0$, we repeat the saddle
point analysis with the full expression (\ref{eq:xit}) for the correlation length. Again,
for intermediate times $t<t_x$, the decay of the average density is given by the
stretched exponential (\ref{eq:stretched}). For times larger than the crossover time
$t_x$ the system realizes that some of the rare regions have developed static order and
contribute to a non-zero long-time limit of the autocorrelation function $C(t)$. The
approach of $C(t)$ to this limiting value is dominated by finite-size islands with
$L_{RR} \sim (-r/A)^{-\nu}$ because they have diverging correlation time.  As a result, we
obtain a power-law,
\begin{equation}
C(t) - C(\infty) \sim t^{-\psi}~ \qquad (\textrm{for}~ t>t_x, ~r<0). \label{eq:power}
\end{equation}
The value of $\psi$ cannot be found by our methods since it depends on the neglected
pre-exponential factors.

In Ref.\ \onlinecite{FendlerSknepnekVojta05}, the predictions (\ref{eq:stretched}),
(\ref{eq:simple}), and (\ref{eq:power}) have been compared to results from a local
dynamical mean-field theory. The purpose of the present paper is to verify these
relations by large-scale kinetic Monte-Carlo simulations of a realistic model with short-range
interactions, viz. the Hamiltonian (\ref{eq:Hamiltonian}).

\section{Computer simulations}
\label{sec:simulations}
\subsection{Method}
\label{subsec:method}

We now turn to the main part of the paper, kinetic Monte-Carlo simulations of our three-dimensional
Ising model with nearest neighbor interactions and planar bond defects, as given in eq.\ (\ref{eq:Hamiltonian}).
Since the smearing of the transition is a result of exponentially rare events, sufficiently large system
sizes are required in order to observe it. In the production runs, we have mostly used system sizes of
$260^3$ or $300^3$ sites. In this context, let us briefly comment on finite-size effects.
Because the phase transition in an Ising model with planar defects is smeared,
conventional finite-size scaling \cite{Barber_review83} does not directly apply.
Nonetheless, there are two types of finite-size effects, whose consequences for the
thermodynamics were discussed in detail in Refs. \onlinecite{Vojta03b,SknepnekVojta04}.
A finite size in the uncorrelated direction limits the possible sizes of the rare regions
and thus cuts off the probability distribution (\ref{eq:w(r)}), while a finite size in
the correlated directions rounds the local phase transitions of the rare regions. We have carried out
test runs using $50^3$, $100^3$ and $200^3$ sites to ensure that our results are not influenced by
these finite-size effects.

We have chosen $J=1$ and $c=0.1$ in eq.\ (\ref{eq:distrib}), i.e., the
strength of a 'weak' bond is 10\% of the strength of a strong bond. The simulations have been performed
for various disorder concentrations $p=\{0.2, 0.3, 0.4, 0.5, 0.6\}$ over a temperature
range of $T=4.25$ to 5.00, covering both sides of the clean critical point at
$T_c^0=4.511$. To suppress the statistical disorder fluctuations we have averaged the
results over a large number of disorder configurations ranging from 100 to 5000 depending
on parameters.

For each disorder configuration, the simulation proceeds as follows: After the
initial setup, the system is equilibrated using the highly efficient Wolff cluster
algorithm.\cite{Wolff89} (Using a cluster algorithm is possible for our Hamiltonian
because the disorder does not induce frustration.) The length of the equilibration
period is 50 Wolff sweeps. (A Wolff sweep is defined as a number of cluster flips
such that the total number of flipped spins is equal to the number of sites,
i.e., on the average each spin is flipped once per sweep.) The actual equilibration
times have typically been of the order of $10$-$20$ sweeps at maximum. Thus, an
equilibration period of 50 sweeps is more than sufficient.

After the system is equilibrated, the spin configuration is stored and the simulation
time is set to $t=0$. The actual time evolution of the system according to model A
dynamics is performed using the Metropolis algorithm.\cite{MRRT53} After each sweep
(a Metropolis sweep consists in one attempted spin flip per site), the spin configuration
is determined and the spin autocorrelation
function is calculated from (\ref{eq:autocorr}). The length of this measurement period is
up to 5000 Monte-Carlo (Metropolis) sweeps.

\subsection{Results}
\label{subsec:results}

\begin{figure}
\includegraphics[width=\columnwidth]{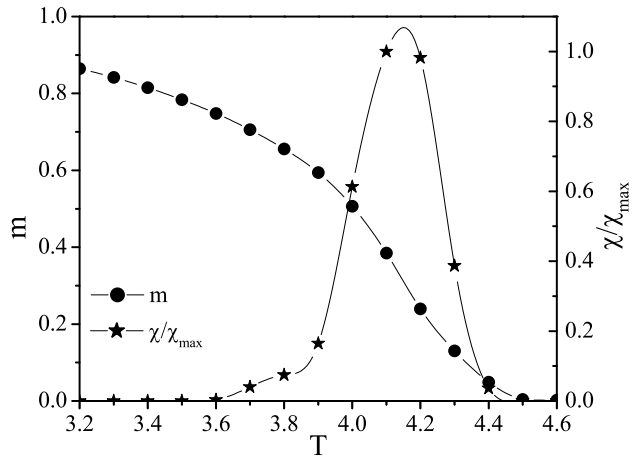}
\caption{Total magnetization $m$ and susceptibility $\chi$ as functions of $T$
for $L_\perp=100$, $L_C=200$ and $p=0.2$ averaged over 200 disorder realizations
(from Ref.\ \onlinecite{SknepnekVojta04}).}
\label{fig:mag}
\end{figure}

In this section, we present the results of our kinetic Monte-Carlo simulations and
compare them to the theoretical predictions of section \ref{subsec:OFT-dynamics}. For
later reference, we first give a brief overview of the thermodynamics, as determined in Ref.\
\onlinecite{SknepnekVojta04}. Fig.\ \ref{fig:mag} shows total magnetization and susceptibility
as functions of temperature averaged over 200 samples of size $L_\perp=100$ and $L_C=200$ with an
impurity concentration $p=0.2$.
For temperatures above $T=4.2 \ldots 4.3$, the magnetization shows a pronounced tail, i.e., it vanishes
very gradually when approaching the clean critical temperature $T_c^0=4.511$. It should
be emphasized that this tail is \emph{not} due to finite-size effects. Indeed, a detailed
analysis\cite{SknepnekVojta04} has shown that the magnetization nicely follows the prediction
(\ref{eq:mtot}) of the smeared transition scenario over more then a magnitude in $m$.
Analogously, an analysis of the susceptibility has shown that it does not diverge in the thermodynamic
limit, instead, it displays a rounded peak.

\begin{figure}
\includegraphics[width=\columnwidth]{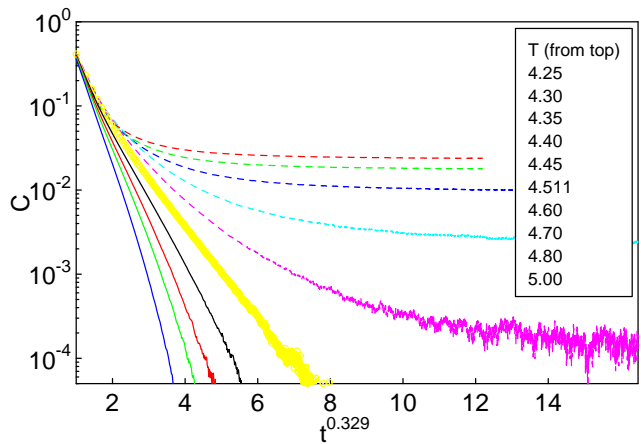}
\caption{Spin autocorrelation function $C$ as a function of time $t$ for $L_\perp=L_C=260$,
$p=0.3$. The number of disorder realizations ranges from 100 at the lowest temperatures to
5000 at the highest temperatures. The power-law time axis corresponds to the stretched
exponential predicted in (\ref{eq:stretched}).  }
\label{fig:evo_stretched}
\end{figure}

We now turn to the spin autocorrelation function  $C(t)$. Figure \ref{fig:evo_stretched} gives an overview
over its behavior for an impurity concentration $p=0.3$ and times up to $t=5000$ over the entire temperature range
studied.
In this figure the data are plotted in the form $\ln C$ vs. $t^{0.329}$ such that the
stretched exponential (\ref{eq:stretched}) predicted for the time dependence at the clean critical
temperature $T_c^0=4.511$ gives a straight line. Here we have used $d_\perp=1$ and a
value of $z=2.04$ for the dynamical exponent of the clean three-dimensional Ising model
with model A dynamics (see, e.g., Refs.\
\onlinecite{WanslebenLandau91,Grassberger95,JMSZ99}). The figure shows that the
autocorrelation function at $T_c^0$ indeed follows the stretched exponential over more
than two orders of magnitude in $C$. At temperatures $T\ne T_c^0$, the autocorrelation
function initially follows the same stretched exponential but eventually crosses over
to a different asymptotic form: For $T>T_c^0$, it decays faster than the stretched exponential,
while it seems to saturate at a nonzero value for $T<T_c^0$. In the
following, we investigate the different temperature regions in more detail.

\begin{figure}
\includegraphics[width=\columnwidth]{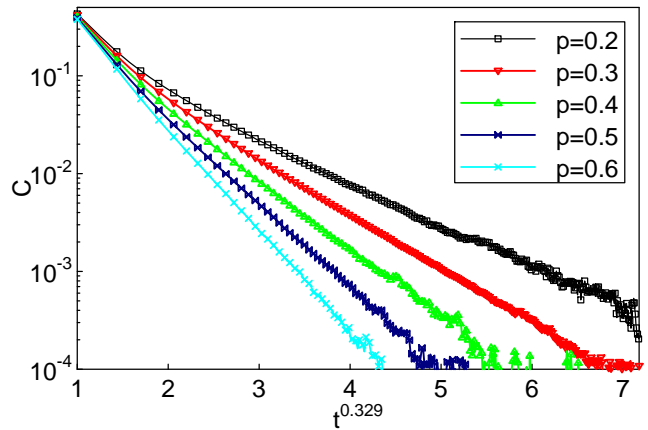}
\caption{Spin autocorrelation function $C(t)$ at the clean critical temperature $T_c^0=4.511$
for $L_\perp=L_C=300$ and several impurity concentrations $p$. The data are averages over 100
disorder realizations (1000 realizations for $p=0.3$). The statistical errors are not bigger than
a symbol size for $C>10^{-3}$. For $p=0.3$, they reach about twice the symbol size at the right end of the curve.}
\label{fig:evo_tc0_compare}
\end{figure}

\begin{figure}
\includegraphics[width=\columnwidth]{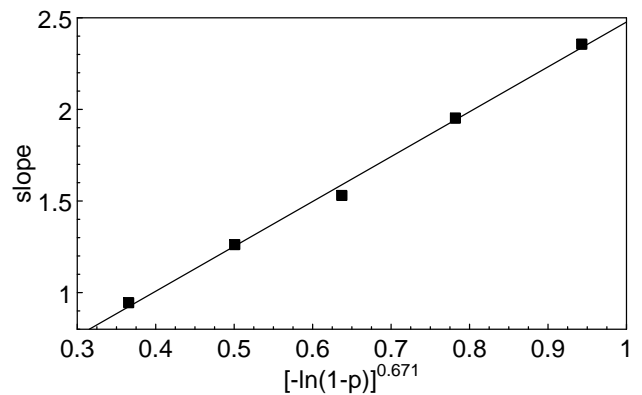}
\caption{Decays constants (slopes) of the stretched exponentials of Fig.\
\ref{fig:evo_tc0_compare} as a function of impurity concentration $p$, plotted according
to the prediction (\ref{eq:stretched}).} \label{fig:slopes_tc0}
\end{figure}

Let us begin by analyzing the autocorrelation function $C(t)$ right at the clean critical
temperature $T_c^0$. Figure \ref{fig:evo_tc0_compare} shows $C(t)$  for several impurity
concentrations $p=0.2 \ldots 0.6$, again plotted as $\ln C$ vs. $t^{0.329}$.
After the initial transients, all curves are straight lines, indicating that $C(t)$
follows the theoretical prediction (\ref{eq:stretched}) over at least two orders of magnitude
in $C$ for all concentrations. From the curves in Fig.\ \ref{fig:evo_tc0_compare}, one
can determine the decay constants (i.e., the slopes) as a function of impurity
concentration $p$. According to (\ref{eq:stretched}), the decay constant should be
proportional to $[-\ln(1-p)]^{z/(d_\bot+z)}=[-\ln(1-p)]^{0.671}$. Figure
\ref{fig:slopes_tc0} shows the corresponding plot for our data.
It shows that the behavior of the decay constants is in good agreement with the
theoretical predictions.

\begin{figure}
\includegraphics[width=\columnwidth]{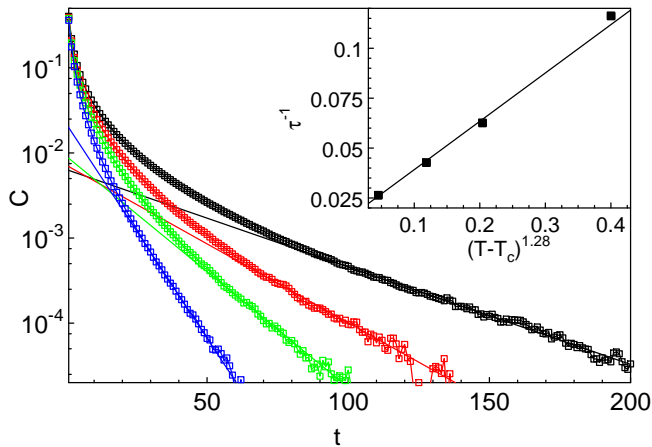}
\caption{Log-linear plot of the spin autocorrelation function for temperatures above $T_c^0$
         (from the top: $T=4.6, 4.7, 4.8, 5.0$).  The other parameters are as in Fig.\
          \ref{fig:evo_stretched}. The solid lines are fits of the long-time behavior to (\ref{eq:simple}).
          The statistical errors are not bigger than a symbol size for $C>10^{-4}$.
          Inset: Inverse decay time $\tau^{-1}$ as a function of $(T-T_c^0)^{z\nu}$.}
\label{fig:evo_above_tc}
\end{figure}

After the behavior at the clean critical temperature $T_c^0$, we now consider the
conventional paramagnetic phase $T>T_c^0$. In Figure \ref{fig:evo_above_tc}, we plot the
autocorrelation function $C(t)$ for $p=0.3, L_\perp=L_C=260$, and temperatures $T>T_c^0$ as
$\ln C$ vs $t$ such that a simple exponential decay gives a straight line.
For all shown temperatures, the long-time behavior of the autocorrelation function is
indeed an exponential decay in agreement with the theoretical prediction
(\ref{eq:simple}). To determine the decay time $\tau$ we fit the long-time behavior of
$C(t)$ to (\ref{eq:simple}). The inset of Fig.\ \ref{fig:evo_above_tc} shows $\tau^{-1}$
as a function of $(T-T_c^0)^{z\nu}$ with the three-dimensional clean correlation length
exponent given by $\nu=0.628$ (see, e.g., Ref.\ \onlinecite{FerrenbergLandau91}) and
$z=2.04$ as before. As predicted by the optimal fluctuation theory in Sec.\
\ref{subsec:OFT-dynamics}, the inverse decay time depends linearly on $(T-T_c^0)^{z\nu}$.
(The remaining small deviations can probably be attributed to the pre-exponential
factors neglected in the optimal fluctuation theory.) We have performed analogous simulations for an impurity
concentration of $p=0.5$ and found the same qualitative behavior. Of course,
non-universal prefactors have different values.

\begin{figure}
\includegraphics[width=\columnwidth]{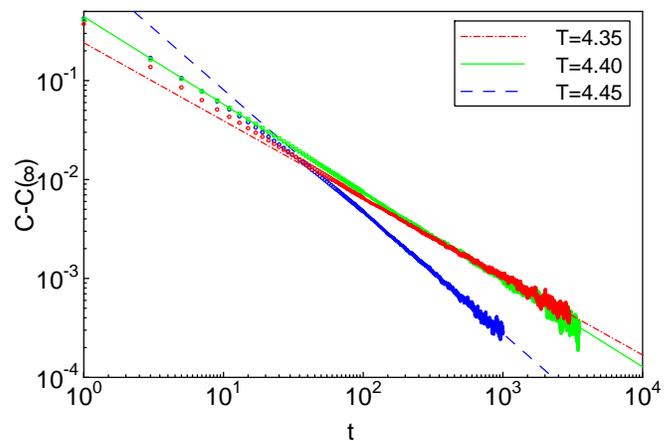}
\caption{Double-logarithmic plot of the dynamical part $C(t)-C(\infty)$ of the autocorrelation function
         for temperatures below $T_c^0$.
         The other parameters are as in Fig.\ \ref{fig:evo_stretched}. The straight lines are fits
         of the long-time behavior to the power law (\ref{eq:power}) giving exponents of $\psi=0.80 ~(T=4.35),~ 0.90
         ~(T=4.40)$ and $1.23 ~(T=4.45)$.}
\label{fig:evo_below_tc}
\end{figure}

Finally, we turn to the properties of the spin autocorrelation function $C(t)$ below the
clean critical temperature $T_c^0$, i.e., in the tail of the smeared phase transition.
For these temperatures, the total equilibrium magnetization is nonzero because
some of the rare regions have already developed a static magnetization (see also
Fig.\ \ref{fig:mag}).\cite{Vojta03b,SknepnekVojta04} Consequently, the autocorrelation
function does not decay to zero in the long-time limit but rather approaches the limiting
value $m^2$ as can be seen in the overview figure \ref{fig:evo_stretched}.
 The dynamical correlations of the magnetization fluctuations $(S_{i,j,k}-m)$
are represented by the deviation $C(t)-C(\infty)$ of the autocorrelation function from
its long-time limit. Figure \ref{fig:evo_below_tc} shows a double-logarithmic plot of $C(t)-C(\infty)$
for $p=0.3, L_\perp=L_C=260$, and temperatures $T=4.35, 4.40$, and 4.45.
For all three temperatures, the long-time behavior of $C(t)-C(\infty)$ follows a power
law. Fits to (\ref{eq:power}) give exponents of $\psi=0.80$ for $T=4.35$, 0.90 for $T=4.40$,
and $1.23$ for ~$T=4.45$. Thus, as predicted, the exponent of the power-law (\ref{eq:power})
is nonuniversal. (We note that the apparent crossing of the curves in Fig.\ \ref{fig:evo_below_tc}
is not related to the long-time dynamics but simply the result of the subtraction of
$C(\infty)$. At lower temperatures, $C(\infty)$ is larger, and since $C(t)$ cannot
exceed 1, the difference $C(t)-C(\infty)$ at early times must decrease with decreasing
$T$. Since in the long-time limit, the decay is faster for larger $T$, the curves must
cross.)

We have carried out analogous simulations for the impurity concentration $p=0.5$, and we
have obtained equivalent results.

\section{Conclusions}
\label{sec:conclusions}

To summarize, we have studied the dynamic behavior of randomly layered Ising magnets
by performing large-scale kinetic Monte-Carlo simulations  of a three-dimensional Ising model
with planar defects. In this system, the magnetic phase transition is smeared  because rare strongly
coupled spatial regions (i.e., thick strongly-coupled layers) independently undergo the phase transition.
We have found that the dynamics in the rare-region dominated tail of the smeared
transition is very slow. The spin autocorrelation function approaches its stationary value
following a power-law in time. At the clean critical temperature $T_c^0$ (which marks the boundary between
the conventional paramagnetic phase and the tail of the smeared transition), the
autocorrelation function decays like a stretched exponential in time. In the
paramagnetic phase above $T_c^0$, the decay is of simple exponential type. However, the
decay time $\tau$ of this exponential diverges in the limit $T\to T_c^0+$. Our numerical
results (both the functional forms and the exponent values of the various time and
disorder dependencies) are in very good agreement with a recent optimal fluctuation
theory.\cite{FendlerSknepnekVojta05}

All our explicit results are for strong impurities ($c=0.1$). We have chosen this value
because strong impurities allow us to easily observe the smeared transition in a
finite-size simulation. If the impurities are weak ($c$ close to 1), the smeared transition
is too close to the clean critical point. Strong bulk fluctuations thus compete with
the smearing, and the latter can only be observed in larger systems and/or at
longer times. The behavior of our model for $c=0$ (dilution) is special, because for this value,
the system is decomposed into an ensemble of noninteracting slabs.

We emphasize that we have considered a purely relaxational (local) dynamics corresponding
to model A in the Hohenberg-Halperin classification. \cite{HohenbergHalperin77} Other dynamic
algorithms require separate investigations. For instance, model B dynamics globally
conserves the order parameter. In this context, an interesting question is: How does the
interplay of the \emph{local} thermodynamics of the rare regions and the \emph{global}
conservation law modify the dynamic behavior in the tail of the smeared transition.

Let us briefly compare the dynamics at a smeared phase transition studied here to the behavior in a
conventional Griffiths phase (as produced by uncorrelated or short-range correlated disorder such
as point defects). Both in a Griffiths phase and at a smeared transition, rare regions dominate the long-time dynamics.
In a conventional Griffiths phase, the finite-size rare regions remain fluctuating for all temperatures
above the dirty critical point, i.e., their island correlation times \emph{remain finite}. As a result, the
autocorrelation function decays like $\ln C(t) \sim -(\ln t)^{d/(d-1)}$ for Ising systems
\cite{Dhar83,RanderiaSethnaPalmer85,DharRanderiaSethna88,Bray88,BrayRodgers88}
and  $\ln C(t) \sim -t^{1/2}$ for Heisenberg systems \cite{Bray88,Bray87}.
In contrast, in the tail of a \emph{smeared} transition, the effects of the rare regions are even stronger
because individual islands can undergo the phase transition independently, connected
with a \emph{divergent} island correlation time. Summing over all islands then leads to an even slower power-law
decay of the spin autocorrelation function.

Let us also comment on experiments. A direct realization of the scenario discussed here
could be achieved by growing alternating layers of two ferromagnetic materials with
different critical temperatures $T_c$. To introduce disorder, the thickness of the layers
should be random. In the tail of the smeared transition, i.e., for temperatures close to
but below the higher of the two $T_c$, the dynamic magnetic response of this system
will be dominated by the rare region contributions. Specifically, the dynamic
susceptibility $\chi(\omega)$ will be dominated by its local part which is essentially
the Fourier transform of the autocorrelation function. Thus measuring the dynamic
response in such a randomly layered magnet should provide an experimental verification of
our results.

Finally, while the results here have been derived for an Ising model with
planar defects, we expect analogous results for other disorder-smeared phase transitions.
Indeed, in the tail of  the smeared non-equilibrium phase transition of a contact process with
extended impurities, a power-law decay of the density was recently found. \cite{Vojta04,DickisonVojta05}.

\begin{acknowledgments}
We acknowledge support from the University of Missouri Research Board, from the
NSF under grants No. DMR-0339147 and PHY99-07949, and from Research Corporation.
Parts of this work have been performed at the Aspen Center for Physics and at the
Kavli Institute for Theoretical Physics, Santa Barbara.
\end{acknowledgments}

\bibliographystyle{apsrev}
\bibliography{../00Bibtex/rareregions}

\begin{thebibliography}{36}
\expandafter\ifx\csname natexlab\endcsname\relax\def\natexlab#1{#1}\fi
\expandafter\ifx\csname bibnamefont\endcsname\relax
  \def\bibnamefont#1{#1}\fi
\expandafter\ifx\csname bibfnamefont\endcsname\relax
  \def\bibfnamefont#1{#1}\fi
\expandafter\ifx\csname citenamefont\endcsname\relax
  \def\citenamefont#1{#1}\fi
\expandafter\ifx\csname url\endcsname\relax
  \def\url#1{\texttt{#1}}\fi
\expandafter\ifx\csname urlprefix\endcsname\relax\def\urlprefix{URL }\fi
\providecommand{\bibinfo}[2]{#2}
\providecommand{\eprint}[2][]{\url{#2}}

\bibitem[{\citenamefont{Griffiths}(1969)}]{Griffiths69}
\bibinfo{author}{\bibfnamefont{R.~B.} \bibnamefont{Griffiths}},
  \bibinfo{journal}{Phys. Rev. Lett.} \textbf{\bibinfo{volume}{23}},
  \bibinfo{pages}{17} (\bibinfo{year}{1969}).

\bibitem[{\citenamefont{Randeria et~al.}(1985)\citenamefont{Randeria, Sethna,
  and Palmer}}]{RanderiaSethnaPalmer85}
\bibinfo{author}{\bibfnamefont{M.}~\bibnamefont{Randeria}},
  \bibinfo{author}{\bibfnamefont{J.~P.} \bibnamefont{Sethna}},
  \bibnamefont{and} \bibinfo{author}{\bibfnamefont{R.~G.}
  \bibnamefont{Palmer}}, \bibinfo{journal}{Phys. Rev. Lett.}
  \textbf{\bibinfo{volume}{54}}, \bibinfo{pages}{1321} (\bibinfo{year}{1985}).

\bibitem[{\citenamefont{Wortis}(1974)}]{Wortis74}
\bibinfo{author}{\bibfnamefont{M.}~\bibnamefont{Wortis}},
  \bibinfo{journal}{Phys. Rev. B} \textbf{\bibinfo{volume}{10}},
  \bibinfo{pages}{4665} (\bibinfo{year}{1974}).

\bibitem[{\citenamefont{Harris}(1975)}]{Harris75}
\bibinfo{author}{\bibfnamefont{A.~B.} \bibnamefont{Harris}},
  \bibinfo{journal}{Phys. Rev. B} \textbf{\bibinfo{volume}{12}},
  \bibinfo{pages}{203} (\bibinfo{year}{1975}).

\bibitem[{\citenamefont{Bray and Huifang}(1989)}]{BrayHuifang89}
\bibinfo{author}{\bibfnamefont{A.~J.} \bibnamefont{Bray}} \bibnamefont{and}
  \bibinfo{author}{\bibfnamefont{D.}~\bibnamefont{Huifang}},
  \bibinfo{journal}{Phys. Rev. B} \textbf{\bibinfo{volume}{40}},
  \bibinfo{pages}{6980} (\bibinfo{year}{1989}).

\bibitem[{\citenamefont{Imry}(1977)}]{Imry77}
\bibinfo{author}{\bibfnamefont{Y.}~\bibnamefont{Imry}}, \bibinfo{journal}{Phys.
  Rev. B} \textbf{\bibinfo{volume}{15}}, \bibinfo{pages}{4448}
  (\bibinfo{year}{1977}).

\bibitem[{\citenamefont{Dhar}(1983)}]{Dhar83}
\bibinfo{author}{\bibfnamefont{D.}~\bibnamefont{Dhar}}, in
  \emph{\bibinfo{booktitle}{Stochastic Processes: Formalism and Applications}},
  edited by \bibinfo{editor}{\bibfnamefont{D.~S.} \bibnamefont{Argawal}}
  \bibnamefont{and}
  \bibinfo{editor}{\bibfnamefont{S.}~\bibnamefont{Dattagupta}}
  (\bibinfo{publisher}{Springer}, \bibinfo{address}{Berlin},
  \bibinfo{year}{1983}).

\bibitem[{\citenamefont{Dhar et~al.}(1988)\citenamefont{Dhar, Randeria, and
  Sethna}}]{DharRanderiaSethna88}
\bibinfo{author}{\bibfnamefont{D.}~\bibnamefont{Dhar}},
  \bibinfo{author}{\bibfnamefont{M.}~\bibnamefont{Randeria}}, \bibnamefont{and}
  \bibinfo{author}{\bibfnamefont{J.~P.} \bibnamefont{Sethna}},
  \bibinfo{journal}{Europhys. Lett.} \textbf{\bibinfo{volume}{5}},
  \bibinfo{pages}{485} (\bibinfo{year}{1988}).

\bibitem[{\citenamefont{Bray}(1988)}]{Bray88}
\bibinfo{author}{\bibfnamefont{A.~J.} \bibnamefont{Bray}},
  \bibinfo{journal}{Phys. Rev. Lett.} \textbf{\bibinfo{volume}{60}},
  \bibinfo{pages}{720} (\bibinfo{year}{1988}).

\bibitem[{\citenamefont{Bray and Rodgers}(1988)}]{BrayRodgers88}
\bibinfo{author}{\bibfnamefont{A.~J.} \bibnamefont{Bray}} \bibnamefont{and}
  \bibinfo{author}{\bibfnamefont{G.~J.} \bibnamefont{Rodgers}},
  \bibinfo{journal}{Phys. Rev. B} \textbf{\bibinfo{volume}{38}},
  \bibinfo{pages}{9252} (\bibinfo{year}{1988}).

\bibitem[{\citenamefont{Bray}(1987)}]{Bray87}
\bibinfo{author}{\bibfnamefont{A.~J.} \bibnamefont{Bray}},
  \bibinfo{journal}{Phys. Rev. Lett.} \textbf{\bibinfo{volume}{59}},
  \bibinfo{pages}{586} (\bibinfo{year}{1987}).

\bibitem[{\citenamefont{v.~Dreyfus et~al.}(1995)\citenamefont{v.~Dreyfus,
  Klein, and Perez.}}]{DreyfusKleinPerez95}
\bibinfo{author}{\bibfnamefont{H.}~\bibnamefont{v.~Dreyfus}},
  \bibinfo{author}{\bibfnamefont{A.}~\bibnamefont{Klein}}, \bibnamefont{and}
  \bibinfo{author}{\bibfnamefont{J.~F.} \bibnamefont{Perez.}},
  \bibinfo{journal}{Commun. Math. Phys.} \textbf{\bibinfo{volume}{170}},
  \bibinfo{pages}{21} (\bibinfo{year}{1995}).

\bibitem[{\citenamefont{Gielis and Maes}(1995)}]{GielisMaes95}
\bibinfo{author}{\bibfnamefont{G.}~\bibnamefont{Gielis}} \bibnamefont{and}
  \bibinfo{author}{\bibfnamefont{C.}~\bibnamefont{Maes}}, \bibinfo{journal}{J.
  Stat. Phys.} \textbf{\bibinfo{volume}{81}}, \bibinfo{pages}{829}
  (\bibinfo{year}{1995}).

\bibitem[{\citenamefont{Cesi et~al.}(1997{\natexlab{a}})\citenamefont{Cesi,
  Maes, and Martinelli}}]{CesiMaesMartinelli97a}
\bibinfo{author}{\bibfnamefont{F.}~\bibnamefont{Cesi}},
  \bibinfo{author}{\bibfnamefont{C.}~\bibnamefont{Maes}}, \bibnamefont{and}
  \bibinfo{author}{\bibfnamefont{F.}~\bibnamefont{Martinelli}},
  \bibinfo{journal}{Commun. Math. Phys.} \textbf{\bibinfo{volume}{188}},
  \bibinfo{pages}{135} (\bibinfo{year}{1997}{\natexlab{a}}).

\bibitem[{\citenamefont{Cesi et~al.}(1997{\natexlab{b}})\citenamefont{Cesi,
  Maes, and Martinelli}}]{CesiMaesMartinelli97b}
\bibinfo{author}{\bibfnamefont{F.}~\bibnamefont{Cesi}},
  \bibinfo{author}{\bibfnamefont{C.}~\bibnamefont{Maes}}, \bibnamefont{and}
  \bibinfo{author}{\bibfnamefont{F.}~\bibnamefont{Martinelli}},
  \bibinfo{journal}{Commun. Math. Phys.} \textbf{\bibinfo{volume}{189}},
  \bibinfo{pages}{323} (\bibinfo{year}{1997}{\natexlab{b}}).

\bibitem[{\citenamefont{McCoy and Wu}(1968{\natexlab{a}})}]{McCoyWu68}
\bibinfo{author}{\bibfnamefont{B.~M.} \bibnamefont{McCoy}} \bibnamefont{and}
  \bibinfo{author}{\bibfnamefont{T.~T.} \bibnamefont{Wu}},
  \bibinfo{journal}{Phys. Rev. Lett.} \textbf{\bibinfo{volume}{21}},
  \bibinfo{pages}{549} (\bibinfo{year}{1968}{\natexlab{a}}).

\bibitem[{\citenamefont{McCoy and Wu}(1968{\natexlab{b}})}]{McCoyWu68a}
\bibinfo{author}{\bibfnamefont{B.~M.} \bibnamefont{McCoy}} \bibnamefont{and}
  \bibinfo{author}{\bibfnamefont{T.~T.} \bibnamefont{Wu}},
  \bibinfo{journal}{Phys. Rev.} \textbf{\bibinfo{volume}{176}},
  \bibinfo{pages}{631} (\bibinfo{year}{1968}{\natexlab{b}}).

\bibitem[{\citenamefont{Fisher}(1992)}]{Fisher92}
\bibinfo{author}{\bibfnamefont{D.~S.} \bibnamefont{Fisher}},
  \bibinfo{journal}{Phys. Rev. Lett.} \textbf{\bibinfo{volume}{69}},
  \bibinfo{pages}{534} (\bibinfo{year}{1992}).

\bibitem[{\citenamefont{Fisher}(1995)}]{Fisher95}
\bibinfo{author}{\bibfnamefont{D.~S.} \bibnamefont{Fisher}},
  \bibinfo{journal}{Phys. Rev. B} \textbf{\bibinfo{volume}{51}},
  \bibinfo{pages}{6411} (\bibinfo{year}{1995}).

\bibitem[{\citenamefont{Vojta}(2003{\natexlab{a}})}]{Vojta03b}
\bibinfo{author}{\bibfnamefont{T.}~\bibnamefont{Vojta}}, \bibinfo{journal}{J.
  Phys. A} \textbf{\bibinfo{volume}{36}}, \bibinfo{pages}{10921}
  (\bibinfo{year}{2003}{\natexlab{a}}).

\bibitem[{\citenamefont{Sknepnek and Vojta}(2004)}]{SknepnekVojta04}
\bibinfo{author}{\bibfnamefont{R.}~\bibnamefont{Sknepnek}} \bibnamefont{and}
  \bibinfo{author}{\bibfnamefont{T.}~\bibnamefont{Vojta}},
  \bibinfo{journal}{Phys. Rev. B} \textbf{\bibinfo{volume}{69}},
  \bibinfo{pages}{174410} (\bibinfo{year}{2004}).

\bibitem[{\citenamefont{Vojta}(2003{\natexlab{b}})}]{Vojta03a}
\bibinfo{author}{\bibfnamefont{T.}~\bibnamefont{Vojta}},
  \bibinfo{journal}{Phys. Rev. Lett.} \textbf{\bibinfo{volume}{90}},
  \bibinfo{pages}{107202} (\bibinfo{year}{2003}{\natexlab{b}}).

\bibitem[{\citenamefont{Vojta}(2004)}]{Vojta04}
\bibinfo{author}{\bibfnamefont{T.}~\bibnamefont{Vojta}},
  \bibinfo{journal}{Phys. Rev. E} \textbf{\bibinfo{volume}{70}},
  \bibinfo{pages}{026108} (\bibinfo{year}{2004}).

\bibitem[{\citenamefont{Vojta}(2006)}]{Vojta06}
\bibinfo{author}{\bibfnamefont{T.}~\bibnamefont{Vojta}}, \bibinfo{journal}{J.
  Phys. A} \textbf{\bibinfo{volume}{39}}, \bibinfo{pages}{R143}
  (\bibinfo{year}{2006}).

\bibitem[{\citenamefont{Fendler et~al.}(2005)\citenamefont{Fendler, Sknepnek,
  and Vojta}}]{FendlerSknepnekVojta05}
\bibinfo{author}{\bibfnamefont{B.}~\bibnamefont{Fendler}},
  \bibinfo{author}{\bibfnamefont{R.}~\bibnamefont{Sknepnek}}, \bibnamefont{and}
  \bibinfo{author}{\bibfnamefont{T.}~\bibnamefont{Vojta}}, \bibinfo{journal}{J.
  Phys. A} \textbf{\bibinfo{volume}{38}}, \bibinfo{pages}{2349}
  (\bibinfo{year}{2005}).

\bibitem[{\citenamefont{Hohenberg and Halperin}(1977)}]{HohenbergHalperin77}
\bibinfo{author}{\bibfnamefont{P.~C.} \bibnamefont{Hohenberg}}
  \bibnamefont{and} \bibinfo{author}{\bibfnamefont{B.~I.}
  \bibnamefont{Halperin}}, \bibinfo{journal}{Rev. Mod. Phys.}
  \textbf{\bibinfo{volume}{49}}, \bibinfo{pages}{435} (\bibinfo{year}{1977}).

\bibitem[{\citenamefont{Metropolis et~al.}(1953)\citenamefont{Metropolis,
  Rosenbluth, Rosenbluth, and Teller}}]{MRRT53}
\bibinfo{author}{\bibfnamefont{N.}~\bibnamefont{Metropolis}},
  \bibinfo{author}{\bibfnamefont{A.}~\bibnamefont{Rosenbluth}},
  \bibinfo{author}{\bibfnamefont{M.}~\bibnamefont{Rosenbluth}},
  \bibnamefont{and} \bibinfo{author}{\bibfnamefont{A.}~\bibnamefont{Teller}},
  \bibinfo{journal}{J. Chem. Phys.} \textbf{\bibinfo{volume}{21}},
  \bibinfo{pages}{1087} (\bibinfo{year}{1953}).

\bibitem[{\citenamefont{Glauber}(1963)}]{Glauber63}
\bibinfo{author}{\bibfnamefont{R.~J.} \bibnamefont{Glauber}},
  \bibinfo{journal}{J. Math. Phys.} \textbf{\bibinfo{volume}{4}},
  \bibinfo{pages}{294} (\bibinfo{year}{1963}).

\bibitem[{\citenamefont{Lifshitz}(1964)}]{Lifshitz64}
\bibinfo{author}{\bibfnamefont{I.~M.} \bibnamefont{Lifshitz}},
  \bibinfo{journal}{Usp. Fiz. Nauk} \textbf{\bibinfo{volume}{83}},
  \bibinfo{pages}{617} (\bibinfo{year}{1964}), \bibinfo{note}{[Sov. Phys.--Usp.
  {\bf 7}, 549 (1965)]}.

\bibitem[{\citenamefont{Barber}(1983)}]{Barber_review83}
\bibinfo{author}{\bibfnamefont{M.~N.} \bibnamefont{Barber}}, in
  \emph{\bibinfo{booktitle}{Phase Transitions and Critical Phenomena}}, edited
  by \bibinfo{editor}{\bibfnamefont{C.}~\bibnamefont{Domb}} \bibnamefont{and}
  \bibinfo{editor}{\bibfnamefont{J.~L.} \bibnamefont{Lebowitz}}
  (\bibinfo{publisher}{Academic}, \bibinfo{address}{New York},
  \bibinfo{year}{1983}), vol.~\bibinfo{volume}{8}, p. \bibinfo{pages}{145}.

\bibitem[{\citenamefont{Wolff}(1989)}]{Wolff89}
\bibinfo{author}{\bibfnamefont{U.}~\bibnamefont{Wolff}},
  \bibinfo{journal}{Phys. Rev. Lett.} \textbf{\bibinfo{volume}{62}},
  \bibinfo{pages}{361} (\bibinfo{year}{1989}).

\bibitem[{\citenamefont{Wansleben and Landau}(1991)}]{WanslebenLandau91}
\bibinfo{author}{\bibfnamefont{S.}~\bibnamefont{Wansleben}} \bibnamefont{and}
  \bibinfo{author}{\bibfnamefont{D.~P.} \bibnamefont{Landau}},
  \bibinfo{journal}{Phys. Rev. B} \textbf{\bibinfo{volume}{43}},
  \bibinfo{pages}{6006} (\bibinfo{year}{1991}).

\bibitem[{\citenamefont{Grassberger}(1995)}]{Grassberger95}
\bibinfo{author}{\bibfnamefont{P.}~\bibnamefont{Grassberger}},
  \bibinfo{journal}{Physica A} \textbf{\bibinfo{volume}{214}},
  \bibinfo{pages}{547} (\bibinfo{year}{1995}).

\bibitem[{\citenamefont{Jaster et~al.}(1999)\citenamefont{Jaster, Mainville,
  Sch{\"u}lke, and Zheng}}]{JMSZ99}
\bibinfo{author}{\bibfnamefont{A.}~\bibnamefont{Jaster}},
  \bibinfo{author}{\bibfnamefont{J.}~\bibnamefont{Mainville}},
  \bibinfo{author}{\bibfnamefont{L.}~\bibnamefont{Sch{\"u}lke}},
  \bibnamefont{and} \bibinfo{author}{\bibfnamefont{B.}~\bibnamefont{Zheng}},
  \bibinfo{journal}{J. Phys. A} \textbf{\bibinfo{volume}{32}},
  \bibinfo{pages}{1395} (\bibinfo{year}{1999}).

\bibitem[{\citenamefont{Ferrenberg and Landau}(1991)}]{FerrenbergLandau91}
\bibinfo{author}{\bibfnamefont{A.~M.} \bibnamefont{Ferrenberg}}
  \bibnamefont{and} \bibinfo{author}{\bibfnamefont{D.~P.}
  \bibnamefont{Landau}}, \bibinfo{journal}{Phys. Rev. B}
  \textbf{\bibinfo{volume}{44}}, \bibinfo{pages}{5081} (\bibinfo{year}{1991}).

\bibitem[{\citenamefont{Dickison and Vojta}(2005)}]{DickisonVojta05}
\bibinfo{author}{\bibfnamefont{M.}~\bibnamefont{Dickison}} \bibnamefont{and}
  \bibinfo{author}{\bibfnamefont{T.}~\bibnamefont{Vojta}}, \bibinfo{journal}{J.
  Phys. A} \textbf{\bibinfo{volume}{38}}, \bibinfo{pages}{1199}
  (\bibinfo{year}{2005}).

\end{thebibliography}

\end{document}